\documentclass[lettersize,journal]{IEEEtran}
\usepackage{amsmath,amsfonts}
\usepackage{algorithmic}
\usepackage{algorithm}
\usepackage{array}
\usepackage[caption=false,font=normalsize,labelfont=sf,textfont=sf]{subfig}
\usepackage{textcomp}
\usepackage{stfloats}
\usepackage{url}
\usepackage{verbatim}
\usepackage{graphicx}
\usepackage{xcolor}
\usepackage{dsfont}
\usepackage{cite}
\usepackage{tikz}
\usetikzlibrary{shapes, arrows.meta, positioning, fit, backgrounds, calc}

\begin{document}

\title{Graph Neural Network-Driven Deep Reinforcement Learning for Scalable RIS Allocation}

\author{\IEEEmembership{Martin Mark Zan}, \IEEEmembership{Stefan Schwarz}, Institute of Telecommunications, TU Wien (Vienna, Austria)
        % <-this % stops a space
\thanks{This research was funded in whole or in part by the
Austrian Science Fund (FWF) 10.55776/PAT4490824. For
open access purposes, the author has applied a CC BY public
copyright license to any author accepted manuscript version
arising from this submission.}% <-this % stops a space
}

% The paper headers
%\markboth{Journal of \LaTeX\ Class Files,~Vol.~14, No.~8, August~2021}%
%{Shell \MakeLowercase{\textit{et al.}}: A Sample Article Using IEEEtran.cls for IEEE Journals}

%\IEEEpubid{0000-0000/00\$00.00~\copyright~2021 IEEE}
% Remember, if you use this you must call \IEEEpubidadjcol in the second
% column for its text to clear the IEEEpubid mark.

\maketitle

% --- IEEE Preprint Notice ---
\begin{tikzpicture}[remember picture, overlay]
  \node[anchor=north, yshift=-1.2cm] at (current page.north) {%
    \parbox{\textwidth}{%
      \centering\footnotesize
      This work has been submitted to the IEEE for possible publication.\\
      Copyright may be transferred without notice, after which this version may no longer be accessible.
    }%
  };
\end{tikzpicture}

\begin{abstract}
Reconfigurable Intelligent Surfaces (RISs) offer a promising paradigm to mitigate blockage and extend millimeter-wave coverage in 6G multi-cell networks. However, dynamically allocating shared RIS infrastructure across competing base stations is an NP-hard problem posing severe scalability bottlenecks. In this paper, we propose a scalable framework combining Graph Neural Networks (GNNs) with Deep Reinforcement Learning (DRL) for dynamic shared RIS orchestration. By formulating allocation as a Markov Decision Process, we introduce a physical topology sparsification strategy that prunes dense channel matrices into a sparse tripartite graph. This pruning reduces edge density by 77\% and removes representation noise, thereby improving global coverage probability while reducing computational complexity. Our relational message-passing architecture naturally generalizes to arbitrary network dimensions without model retraining. Furthermore, structural ablation studies reveal that physical path loss localizes surface dependencies, enabling a highly efficient localized graph design with linear computational scaling. Extensive simulations in dense urban environments demonstrate that under strict infrastructure budget constraints, the proposed GNN-DRL framework consistently outperforms greedy heuristic baselines by up to $\sim$ 12\% in coverage while delivering faster inference speed via GPU acceleration.
\end{abstract}


\section{Introduction}
\label{sec:introduction}

Demand for ubiquitous, high-capacity wireless connectivity has pushed next-generation 6G networks toward millimeter-wave (mmWave) and terahertz (THz) bands. While these bands unlock massive bandwidth, they suffer from severe propagation attenuation and are highly vulnerable to physical blockages, creating significant coverage holes in dense urban environments \cite{Zeng2021, Toumi2021}. To overcome these line-of-sight (LOS) obstructions, Reconfigurable Intelligent Surfaces (RISs) have emerged as a core paradigm-shifting technology. By deploying cost-effective, passive arrays of reflecting elements, an RIS can dynamically manipulate the propagation environment, establishing virtual, high-gain cascaded paths to extend network coverage to previously unreachable cell-edge user equipments (UEs) \cite{Mosleh2023, Le2021}.

Rather than dedicating private, costly reflecting hardware to individual cells, future deployments may favor treating RISs as a shared infrastructure layer. Under this model, an independent operator dynamically leases or assigns a pool of shared RISs to multiple competing base stations (BSs) to maximize global network utility and coverage \cite{Schwarz2024}, or coordinates across multi-operator resource optimization frameworks \cite{Zhang2025}. However, dynamically orchestrating the assignment of these shared surfaces is an NP-hard combinatorial optimization problem. Traditional mathematical programming frameworks scale exponentially with the network size, whereas cooperative multi-cell coordination schemes-such as Coordinated Multi-Point (CoMP) \cite{Irram2020, Schwarz2014} or user-centric cell-free massive MIMO architectures \cite{Nayebi2015, Ammar2022}-and greedy heuristics scale polynomially. In cell-free networks, coordinating large-scale joint transmissions typically relies on decentralized user clustering or specialized learning paradigms \cite{Mendoza2026, Mendoza2023}. As the density of base stations, users, and reconfigurable surfaces scales in massive 6G layouts, recursively computing dense, network-wide cascaded channel matrices to evaluate greedy choices becomes a prohibitive computational bottleneck, rendering traditional optimization entirely impractical for real-time operations.

To address these resource allocation hurdles, market-driven and game-theoretic protocols have recently been explored. In our previous works \cite{Zan2026a, Zan2026b}, we investigated decentralized, auction-based frameworks to orchestrate shared RIS allocation. These methods are deeply rooted in algorithmic mechanism design \cite{Nisan2001} and game-theoretic resource orchestration \cite{Roughgarden2016}, which have historically governed spectrum allocation \cite{Cramton2017} and utility bidding under budget constraints \cite{Milgrom2004}. Specifically, in \cite{Zan2026a}, we designed a simultaneously ascending auction format where competing BSs acted as self-interested deep reinforcement learning (DRL) agents bidding for local RIS units under rental budgets, utilizing reinforcement learning to model market power and strategic bidding under uncertainty \cite{Nanduri2007}. In \cite{Zan2026b}, we extended this paradigm to a collaborative, multi-agent bidding framework to enforce service fairness across highly unbalanced multi-cell environments. However, while these auction-based schemes offer excellent modularity and protect base station privacy, they require iterative bidding rounds and rely on decentralized heuristic utility estimations. These protocol-level overheads limit their execution speed and optimization capacity when scaling up to extremely dense, large-scale networks.

To overcome the scalability limits of decentralized bidding protocols, we shift our focus in this work toward a centralized, end-to-end framework utilizing Graph Neural Networks (GNNs) \cite{Schlichtkrull2017, Corso2020}. Standard DRL agents rely on fixed-input architectures such as Multi-Layer Perceptrons (MLPs) or Convolutional Neural Networks that fail to capture the topological relationships of wireless links and cannot generalize to variations in network size, requiring complete retraining if nodes are added or removed. GNNs, by contrast, model the network topology as a graph and utilize localized message-passing operations to extract scale-invariant node and edge embeddings \cite{Schlichtkrull2017, Corso2020}. By integrating a heterogeneous GNN with a centralized Advantage Actor-Critic (A2C) policy \cite{Mnih2016}, we can construct an allocation engine that naturally generalizes to arbitrary wireless geometries and layout sizes.

A major challenge in deploying GNNs for ultra-dense wireless topologies is the computational overhead of message propagation across massive, fully connected channel graphs. To resolve this, we propose an aggressive physical graph sparsification strategy that prunes links lacking geometric LOS viability or those experiencing severe path loss. This results in a highly sparse, tripartite graph representing the physical environment, allowing the GNN to focus its learning capacity strictly on propagation paths capable of yielding constructive beamforming gains. Furthermore, while global coordination mechanisms such as Self-Attention \cite{Vaswani2017} are often proposed to resolve submodular resource conflicts, we conduct a detailed structural ablation study. We demonstrate that physical path loss fundamentally localizes RIS coverage dependencies, meaning that a purely localized graph convolutional architecture can achieve identical performance to attention-based models while reducing computational complexity.

In this paper, we propose a scalable, heterogeneous GNN-driven DRL framework to solve the dynamic allocation of shared RIS infrastructure in large-scale multi-cell networks. The major contributions of this work are summarized as follows:
\begin{itemize}
    \item \textbf{Heterogeneous GNN-DRL Formulation:} We formulate the multi-cell shared RIS allocation problem as an episodic, sequential Markov Decision Process (MDP) and develop an Advantage Actor-Critic learning framework powered by a multi-relational GNN. This architecture natively processes arbitrary network dimensions and accounts for the distinct physical identities of BSs, RISs, and users.
    \item \textbf{Physical Graph Sparsification:} We design a topology-pruning strategy based on physical LOS visibility and comparative path-loss offsets. This reduces the dense propagation channel into a highly sparse tripartite graph (yielding a $77\%$ total edge reduction) while simultaneously improving global coverage performance and training efficiency over dense graph baselines.
    \item \textbf{Asymptotic Complexity Reduction:} We prove that our localized message-passing framework scales strictly linearly, $\mathcal{O}(K \cdot L(|\mathcal{V}| + |\mathcal{E}|))$, with respect to the sparsified graph size. Through detailed ablation studies, we demonstrate that a global multi-head self-attention layer is unnecessary, empirically validating that physical path loss localizes allocation dependencies and justifies a highly scalable, attention-free graph convolutional design.
    \item \textbf{Numerical Evaluation and Baseline Comparisons:} 
    We validate the proposed model through extensive large-scale simulations. 
    Our results demonstrate that when active infrastructure resources are heavily constrained, the GNN-DRL framework consistently outperforms greedy heuristic baselines by up to 12\%, delivering superior global coverage.
\end{itemize}

The rest of this paper is structured as follows: Section \ref{sec:system_model} presents the propagation, signal, and problem formulations. Section \ref{sec:gnn_architecture} outlines the MDP and the proposed GNN-DRL architecture. Section \ref{sec:complexity} evaluates the computational complexities. Sections V and VI discuss the ablation studies and the numerical results, respectively, before the paper concludes in Section VII.

\section{System Model and Problem Formulation}
\label{sec:system_model}

We consider a multi-cell downlink wireless network operating in a highly dense urban environment. The network consists of a set of base stations denoted by $\mathcal{B}$, a set of shared reconfigurable intelligent surfaces denoted by $\mathcal{R}$, and a set of user equipments denoted by $\mathcal{U}$. The RISs are deployed as dynamically shared infrastructure to enhance coverage, particularly at the cell edges where the direct line-of-sight is frequently blocked.

\subsection{Large-Scale Propagation Model}
The signal power attenuation over a communication link between any two nodes in the network is governed by a distance-dependent large-scale path loss function combined with shadow fading.

A communication link to a user $u \in \mathcal{U}$ can be established either directly from a base station $b \in \mathcal{B}$ or via a reflecting surface $r \in \mathcal{R}$.

When an RIS is utilized, the signal travels along a cascaded path from the BS to the RIS, and subsequently from the RIS to the UE. Assuming the RIS elements are coherently aligned to maximize the signal power toward the target user, the effective power gain introduced by the RIS is proportional to the square of its reflecting elements, $M_{\text{RIS}}$ \cite{Balanis2005}:
\begin{equation}
\label{eq:ris_gain}
G_{\text{RIS}} = C_{\text{cluster}} \cdot M_{\text{RIS}}^2,
\end{equation}
where $C_{\text{cluster}} \in (0, 1]$ is a proportionality constant reflecting the hardware array efficiency, and deployment losses.

A communication link between two nodes can operate in either a Line-of-Sight (LOS) or Non-Line-of-Sight (NLOS) state, with path loss exponents and shadow fading variances conditioned on this binary propagation status. The baseline probability of establishing an unobstructed LOS link decreases exponentially with distance $d$:
\begin{equation}
\label{eq:los_prob}
p_{\text{LOS}}(d) = \exp\left(-\frac{d}{d_{\text{LOS, type}}}\right),
\end{equation}
where $d_{\text{LOS, type}}$ is a link-type-dependent scaling parameter (e.g., BS-to-UE, BS-to-RIS, or RIS-to-UE). 
To ensure realistic spatial consistency across the network-such that closely located links share similar environmental blockages-the LOS probabilities are mapped to thresholds over a spatially correlated Gaussian random map~\cite{Schwarz2016_LOS}. The algorithmic generation of these blockage maps and their specific numerical parameters are part of Section~\ref{sec:numerical_results}.

\subsection{Signal and SINR Model}
Let $\gamma_{u,b}$ denote the direct path gain between BS $b$ and UE $u$. Similarly, let $\gamma_{r,b}$ and $\gamma_{u,r}$ denote the unamplified large-scale attenuation gains of the BS-to-RIS link $(b,r)$ and the RIS-to-UE link $(r,u)$, respectively.

The assignment of RISs to base stations is captured by a binary decision matrix $\mathbf{X} \in \{0, 1\}^{|\mathcal{R}| \times |\mathcal{B}|}$, where $x_{r,b} = 1$ if RIS $r$ is assigned to serve BS $b$, and $x_{r,b} = 0$ otherwise. An RIS can be assigned to at most one BS at any given time.

Physical RIS array gain relies on real-time phase alignment tuned to a specific base station's transmission. When an RIS is unassigned ($x_{r,b} = 0$), its reflecting elements remain uncoordinated, causing incident signals to scatter diffusely without coherent beamforming gain, effectively acting as an ordinary wall surface. Consequently, treating uncoordinated scattered reflections as negligible background noise, the total effective channel gain $\gamma_{u,b}^{\text{total}}$ from BS $b$ to UE $u$ combines the direct path with the coherent reflections from active assigned surfaces:
\begin{equation}
\label{eq:total_gain}
\gamma_{u,b}^{\text{total}} = \gamma_{u,b} + \sum_{r \in \mathcal{R}} x_{r,b} \cdot G_{\text{RIS}} \cdot \gamma_{u,r} \cdot \gamma_{r,b}.
\end{equation}

Let $P_b$ denote the transmit power of base station $b$, the received signal power at user $u$ from a given base station $b$ is $P_{u,b} = P_b \gamma_{u,b}^{\text{total}}$. Users dynamically associate with the base station providing the strongest received signal. Thus, the effective received power for user $u$ is:
\begin{equation}
\label{eq:received_power}
P_u = \max_{b \in \mathcal{B}} \left( P_{u,b} \right).
\end{equation}

The corresponding Signal-to-Noise Ratio (SNR) for user $u$ is:
\begin{equation}
\label{eq:user_snr}
\text{SNR}_u = \frac{P_u}{\sigma_n^2},
\end{equation}
where $\sigma_n^2$ is the noise power.

\subsection{Problem Formulation}
The primary objective of the network operator is to maximize the global coverage probability, defined as the ratio of users whose received SNR exceeds a predefined quality-of-service threshold $\text{SNR}_{\min}$. The operator is constrained by a maximum active RIS budget $K$, representing the operational or leasing limits of the shared infrastructure. 

The discrete RIS assignment problem can be formulated as follows:
\begin{subequations}\label{eq:optimization}
\begin{align}
\max_{\mathbf{X}} \quad & \frac{1}{|\mathcal{U}|} \sum_{u \in \mathcal{U}} \mathbb{I}(\text{SNR}_u \geq \text{SNR}_{\min}) \label{eq:opt_obj} \\
\text{s.t.} \quad & \sum_{r \in \mathcal{R}} \sum_{b \in \mathcal{B}} x_{r,b} \leq K, \label{eq:opt_budget} \\
& \sum_{b \in \mathcal{B}} x_{r,b} \leq 1, \quad \forall r \in \mathcal{R}, \label{eq:opt_unique} \\
& x_{r,b} \in \{0, 1\}, \quad \forall r \in \mathcal{R}, \forall b \in \mathcal{B}, \label{eq:opt_binary}
\end{align}
\end{subequations}
where $\mathbb{I}(\cdot)$ is the indicator function. Constraint \eqref{eq:opt_budget} enforces the maximum RIS activation budget, while \eqref{eq:opt_unique} ensures that an RIS is exclusively assigned to a single base station. Note that this objective function can be generalized to account for distinct Quality-of-Service (QoS) requirements across different user groups by replacing $\text{SNR}_{\min}$ with user-specific target thresholds $\text{SNR}_{\min, u}$. This budget constraint implicitly serves as an operational proxy for surface leasing costs.

\section{Proposed Deep RL and GNN Architecture}
\label{sec:gnn_architecture}

To solve the NP-hard combinatorial problem formulated in Section~\ref{sec:system_model}, we design a scalable framework that blends deep reinforcement learning with a heterogeneous graph neural network. Solvers based on classical optimization-including continuous relaxations of binary assignment constraints ($x_{r,b} \in [0, 1]$)-struggle with this formulation because the objective function contains step-like discrete indicators $\mathbb{I}(\cdot)$ and non-smooth max-associations $\max_b(\cdot)$. These constructs induce zero-gradient plateaus and severe non-convexities, leading to poor convergence for gradient-based continuous solvers and substantial optimality gaps upon rounding~\cite{Boyd2004}. Moreover, re-solving numerical optimization programs from scratch for every topological update imposes prohibitive latency in real-time 6G scheduling. To achieve low-latency execution, we recast the assignment problem as an episodic, finite-horizon Markov Decision Process (MDP), allowing a learning agent to construct the assignment matrix $\mathbf{X}$ sequentially.

\subsection{Markov Decision Process Formulation}
\label{subsec:markov}
The MDP is defined by the tuple $(\mathcal{S}, \mathcal{A}, \mathcal{P}, \mathcal{R}, \gamma)$, denoting the state space, action space, transition probability distribution, reward function, and discount factor. 

\subsubsection{State Space $\mathcal{S}$}
At any sequential decision step $t$, the state $s_t \in \mathcal{S}$ must completely encapsulate the network conditions to preserve the Markov property. We formally define the environment state at step $t$ as a composite tuple $s_t = (\mathbf{X}_t, G_t)$. Here, $\mathbf{X}_t \in \{0, 1\}^{|\mathcal{R}| \times |\mathcal{B}|}$ is the partial RIS assignment matrix tracking all historical resource allocations up to step $t$. 

The second component, $G_t = (\mathcal{V}, \mathcal{E}, \mathbf{F}^{(t)})$, represents the physical network structure as a heterogeneous tripartite graph. Here, $\mathcal{V} = \mathcal{B} \cup \mathcal{R} \cup \mathcal{U}$ and $\mathcal{E}$ define the fixed physical graph topology, while $\mathbf{F}^{(t)} = \{\mathbf{f}_i^{(t)}\}_{i \in \mathcal{V}}$ collects the node feature vectors at decision step $t$. The node features are parameterized as follows:
\begin{itemize}
    \item \textbf{BS nodes ($b \in \mathcal{B}$):} $\mathbf{f}_b = [X, Y, P_b]$, containing the static normalized spatial location and transmit power of the base station ($\mathbf{f}_b^{(t)} \equiv \mathbf{f}_b, \forall t$).
    \item \textbf{RIS nodes ($r \in \mathcal{R}$):} $\mathbf{f}_r^{(t)} = [X, Y, K - t, d_r^{\mathcal{U}}, d_r^{\mathcal{B}}]$, representing the normalized coordinates, the remaining activation budget at step $t$, and the node degrees with respect to users and base stations, respectively.
    \item \textbf{User nodes ($u \in \mathcal{U}$):} $\mathbf{f}_u^{(t)} = [X, Y, \text{SNR}_{\min}, \mathds{1}_u^{(t)}]$, containing the normalized location of the user, the minimum required SNR threshold, and a dynamic binary indicator $\mathds{1}_u^{(t)}$ which equals 1 if the user is currently covered, and 0 otherwise.
\end{itemize}

Including 2D spatial coordinates $[X,Y]$ provides explicit geometric layout awareness to the graph neural network. For base stations and reconfigurable surfaces, spatial coordinates represent static infrastructure parameters known upon deployment. For user equipments, accurate real-time location estimates are expected to be readily supplied by emerging 6G Integrated Sensing and Communication (ISAC) frameworks and cellular positioning protocols~\cite{Liu2022_ISAC}. While the edge attributes capture large-scale link attenuations (including random shadow fading), the node-level coordinate features complement these edge weights by supplying global spatial geometry, enabling message-passing layers to recognize user clustering patterns and geometric proximity independently of local shadowing fluctuations.

The edge set $\mathcal{E} = \mathcal{E}_{\mathcal{B}\mathcal{U}} \cup \mathcal{E}_{\mathcal{B}\mathcal{R}} \cup \mathcal{E}_{\mathcal{R}\mathcal{U}}$ encompasses three classes of undirected links weighted by their respective symmetric, large-scale path loss values in dB: $\text{PL}_{b,u}$, $\text{PL}_{b,r}$, and $\text{PL}_{r,u}$. To focus the computational capacity of the GNN on candidate allocations that are both physically viable and potentially beneficial for user coverage, we implement a strict graph sparsification strategy:
\begin{enumerate}
    \item \textbf{Physical Line-of-Sight Viability:} We exclude any base station-to-RIS link $(b,r)$ or RIS-to-user link $(r,u)$ that lacks a clear geometric LOS path. This directly aligns with the propagation constraints established in Section~\ref{sec:system_model}, where cascaded smart reflections require coherent dual-LOS visibility to yield an $M_{\text{RIS}}^2$ beamforming gain.
    \item \textbf{Performance-Based Pruning:} To limit graph complexity and reduce message-passing noise, links whose path losses exceed a small threshold offset from the best available alternative link for a given node are pruned. For direct BS-UE links, this pruning window is bounded by $\Delta_{\text{BU}}$, and for RIS-UE links, it is bounded by $\Delta_{\text{RU}}$. While selecting excessively restrictive thresholds ($\Delta_{\text{BU}}, \Delta_{\text{RU}} \to 0\text{ dB}$) risks discarding viable candidate pathways, moderate thresholding strips out non-contributing channels without incurring suboptimality.
\end{enumerate}
This deliberate pruning strategy yields a highly efficient, sparse topological graph that omits baseline channels incapable of contributing significantly to coverage improvement. While the GNN processes only the sparsified structural graph $G_t$ to maximize scalability, the simulator state $s_t$ tracks $\mathbf{X}_t$ for deterministic reward evaluation.

\subsubsection{Action Space $\mathcal{A}$}
\label{subsubsec:action_space}
Predicting the full binary assignment matrix $\mathbf{X}$ in a single step induces an exponentially large combinatorial action space of size $\mathcal{O}(|\mathcal{B}|^K)$, which severely hinders policy exploration and convergence. To circumvent this, an action $a_t \in \mathcal{A}$ corresponds to selecting a single unassigned RIS $r \in \mathcal{R}$ and mapping it to a specific base station candidate $b \in \mathcal{B}$ at step $t$, decomposing the task into $K$ manageable sequential allocations. Although the SNR indicator threshold in \eqref{eq:opt_obj} is not strictly submodular due to localized step-like gains, the macroscopic accumulation of cascaded channel power exhibits diminishing marginal returns. This weak submodularity theoretically motivates our sequential MDP structure and explains why step-by-step allocations achieve strong optimization performance.

To ensure compliance with the execution criteria defined in \eqref{eq:opt_unique}, the valid action space is restricted dynamically at each decision step via a policy mask. A candidate pair $(r, b)$ is selectable if and only if:
\begin{enumerate}
    \item RIS $r$ has not been activated or assigned in a previous operational step $t' < t$.
    \item An edge connection $(b,r)$ exists within the sparsified structural edge set $\mathcal{E}_{\mathcal{B}\mathcal{R}}$ defined in the graph topology state.
\end{enumerate}
This restricts exploration strictly to physically valid and high-utility propagation pathways. The sequential assignment episode terminates immediately when the step count reaches the terminal resource horizon $t = K$.

\subsubsection{Transition Probability $\mathcal{P}$}
Since the node positions and large-scale path loss parameters remain invariant over the optimization horizon, environmental state transitions are strictly deterministic. Taking action $a_t = (r, b)$ shifts the dynamic tracking parameters by marking RIS $r$ as unavailable, decrementing the operational resource scalar, and updating the downstream user signal states based on \eqref{eq:total_gain}. While this formulation focuses on quasi-static propagation conditions during episodic resource orchestration, extending the MDP framework to dynamic scenarios with user mobility and time-varying channel blockages represents a compelling direction for future work~\cite{Mendoza2026}.

\subsubsection{Reward Design $\mathcal{R}$}
Evaluating training gradients directly against a discontinuous indicator function (as seen in \eqref{eq:opt_obj}) leads to sparse reward bottlenecks. To supply the reinforcement learning agent with rich, step-by-step gradient paths, we employ a continuous proxy function based on a logistic sigmoid mapping that provides a differentiable surrogate for coverage. For each user $u \in \mathcal{U}$, we define a soft coverage measure using the standard sigmoid function:
\begin{equation}
\label{eq:soft_coverage}
\tilde{c}_u(t) = \sigma \left( \kappa \cdot \frac{\text{SNR}_u(t) - \text{SNR}_{\min}}{\text{SNR}_{\min}} \right),
\end{equation}
where $\sigma(z) = 1 / (1 + \exp(-z))$, $\text{SNR}_u(t)$ is the current user SNR calculated using \eqref{eq:user_snr}, and $\kappa$ is a tunable parameter controlling the steepness of the smooth transition. 

The overall network state utility is then defined as the average soft coverage probability across all users:
\begin{equation}
\label{eq:smooth_reward_eval}
\tilde{R}(\mathbf{X}_t) = \frac{1}{|\mathcal{U}|} \sum_{u \in \mathcal{U}} \tilde{c}_u(t).
\end{equation}

To align optimization incentives with sequential progress, the step reward tracking metric provided to the agent is formulated as the dynamic marginal utility gain achieved by the action $a_t$:
\begin{equation}
\label{eq:marginal_reward}
R_t = \tilde{R}(\mathbf{X}_t) - \tilde{R}(\mathbf{X}_{t-1}).
\end{equation}
By maximizing the cumulative sum of these dynamic step rewards $\sum_{t=1}^K R_t$, the centralized policy intrinsically maximizes the absolute global coverage probability at the terminal horizon.

To effectively learn the sequential allocation policy parameterized by this MDP structure, we employ the Advantage Actor-Critic reinforcement learning algorithm \cite{Mnih2016}. The centralized architecture utilizes a deep heterogeneous network to simultaneously approximate both the action allocation policy distribution (the Actor) and the state-value function tracking expected cumulative returns (the Critic). This algorithmic split ensures stable gradient propagation over large action spaces and allows the policy to guide allocation using predictive value estimates. The precise architectural layers mapping the heterogeneous state observations onto these respective Actor-Critic prediction heads are developed in the remaining subsections.

\subsection{Heterogeneous Graph Topology and Input Embedding}
\label{subsec:graph_construction}

We define the set of node types as $\mathcal{T}_v = \{\text{BS}, \text{RIS}, \text{UE}\}$. To capture the directional flow of message passing across the undirected physical channels, the set of canonical edge relation types is defined as a collection of directed node-type pairs:
\begin{equation}
\begin{split}
\mathcal{T}_e = \{ &(\text{BS}, \text{UE}), (\text{UE}, \text{BS}), (\text{BS}, \text{RIS}), \\ &(\text{RIS}, \text{BS}), (\text{RIS}, \text{UE}), (\text{UE}, \text{RIS}) \},
\end{split}
\end{equation}
where each relation $r = (\alpha, \beta) \in \mathcal{T}_e$ indexes one of the corresponding physical link classes ($\mathcal{E}_{\mathcal{B}\mathcal{U}}$, $\mathcal{E}_{\mathcal{B}\mathcal{R}}$, or $\mathcal{E}_{\mathcal{R}\mathcal{U}}$) directionally. While the physical path loss attributes remain symmetric ($\text{PL}_{i,j} = \text{PL}_{j,i}$), this configuration ensures that downstream message-passing functions are unique, independent, and strictly conditioned on the directional identity of the node-type interaction.

The raw feature vectors $\mathbf{f}_b$, $\mathbf{f}_r^{(t)}$, and $\mathbf{f}_u^{(t)}$ possess distinct dimensionalities ($d_{\text{BS}} = 3$, $d_{\text{RIS}} = 5$, and $d_{\text{UE}} = 4$). To eliminate scale disparities and establish a uniform hidden dimension $d_{\text{hidden}}$ for graph convolutions, we introduce a dedicated bank of input linear embedding layers. 

Let $\mathcal{V}_{\tau} \subset \mathcal{V}$ denote the subset of nodes belonging to type $\tau \in \mathcal{T}_v$ (where $\mathcal{V}_{\text{BS}} = \mathcal{B}$, $\mathcal{V}_{\text{RIS}} = \mathcal{R}$, and $\mathcal{V}_{\text{UE}} = \mathcal{U}$). For any node $i \in \mathcal{V}_{\tau}$, the initial latent embedding vector $\mathbf{h}_{i}^{(0)} \in \mathbb{R}^{d_{\text{hidden}}}$ is computed via a node-type-specific linear projection header:
\begin{equation}
\label{eq:linear_embedding}
\mathbf{h}_{i}^{(0)} = \text{ReLU}\left( \mathbf{W}_{\tau}^T \mathbf{f}_{i}^{(t)} + \mathbf{b}_{\tau} \right), \quad \forall i \in \mathcal{V}_{\tau}, \, \forall \tau \in \mathcal{T}_v,
\end{equation}
where $\mathbf{W}_{\tau} \in \mathbb{R}^{d_{\tau} \times d_{\text{hidden}}}$ and $\mathbf{b}_{\tau} \in \mathbb{R}^{d_{\text{hidden}}}$ represent the trainable projection weight matrix and bias vector unique to node type $\tau$. This compact transformation maps the heterogeneous physical network characteristics into dense, uncoordinated hidden representations, providing the initial node states for relation-aware message passing.

\subsection{Heterogeneous Message Passing Layer}
\label{subsec:gnn_message_passing}

To update the node latent embeddings over the sparsified structural graph topology, we construct a multi-layer relational message passing framework inspired by Relational Graph Convolutional Networks (R-GCNs)~\cite{Schlichtkrull2017}. Specifically, we develop a custom multi-relation convolution operator, termed \texttt{WeightedEdgeConv}. Let $l \in \{1, \dots, L\}$ denote the graph neural network layer index. At each layer, information is propagated across all canonical edge relation types $r \in \mathcal{T}_e$ simultaneously, conditioning the structural embeddings on both neighbor node states and localized propagation paths.

\subsubsection{Relational Message and Dual Aggregation}
Consider a specific directional relation $r = (\alpha,\beta) \in \mathcal{T}_e$, which maps connections from a source node type $\alpha \in \mathcal{T}_v$ to a target node type $\beta \in \mathcal{T}_v$, with corresponding node subsets $\mathcal{V}_{\alpha}, \mathcal{V}_{\beta} \subset \mathcal{V}$. For a given target node $i \in \mathcal{V}_{\beta}$, let $\mathcal{N}_r(i) \subseteq \mathcal{V}_{\alpha}$ represent its set of incoming neighbors under relation $r$. The message propagated from a source neighbor node $j \in \mathcal{N}_r(i)$ incorporates the physical link attenuation by concatenating the source's latent feature vector $\mathbf{h}_j^{(l-1)} \in \mathbb{R}^{d_{\text{hidden}}}$ with the scalar path loss feature $\text{PL}_{i,j}$:
\begin{equation}
\label{eq:message_func}
\mathbf{m}_{i,j}^{(r)} = \text{MLP}_{\text{msg}}^{(r)} \left( \left[ \mathbf{h}_j^{(l-1)} \,\|\, \text{PL}_{i,j} \right] \right),
\end{equation}
where $\|$ denotes the vector concatenation operator, and $\text{MLP}_{\text{msg}}^{(r)}: \mathbb{R}^{d_{\text{hidden}} + 1} \to \mathbb{R}^{d_{\text{hidden}}}$ is a relation-specific multi-layer perceptron comprising two linear transformations separated by a Rectified Linear Unit ($\text{ReLU}$) activation.

To robustly capture both localized signal variations and macroscopic topological trends, the layer computes two distinct spatial aggregation metrics across the target node's neighborhood\cite{Corso2020}: the element-wise mean and the element-wise maximum. The dual-aggregated message representation for target node $i$ under relation $r$ is constructed via a linear projection of their concatenation:
\begin{equation}
\label{eq:aggregation_func}
\mathbf{\bar{m}}_i^{(r)} = \mathbf{W}_{\text{aggr}}^{(r)} \left[ \frac{1}{|\mathcal{N}_r(i)|}\sum_{j \in \mathcal{N}_r(i)} \mathbf{m}_{i,j}^{(r)} \;\Big\|\; \max_{j \in \mathcal{N}_r(i)} \left( \mathbf{m}_{i,j}^{(r)} \right) \right],
\end{equation}
where $\mathbf{W}_{\text{aggr}}^{(r)} \in \mathbb{R}^{d_{\text{hidden}} \times 2d_{\text{hidden}}}$ is a trainable projection matrix unique to the relation type.

\subsubsection{Multi-Relation Integration and Layer Updates}
Because a single node type $\beta \in \mathcal{T}_v$ simultaneously experiences interactions from multiple incoming edge classes, the total relational message is accumulated across all active incoming relations. The target node's representation is subsequently regularized by adding its own historical state via a self-loop transformation (root connection). The intermediate update for node $i \in \mathcal{V}_{\beta}$ is formulated as:
\begin{equation}
\label{eq:multi_relation_update}
\mathbf{\tilde{h}}_i^{(l)} = \sum_{r \in \mathcal{T}_e \mid \text{tgt}(r)=\beta} \left( \mathbf{\bar{m}}_i^{(r)} + \mathbf{W}_{\text{root}}^{(r)} \mathbf{h}_i^{(l-1)} \right),
\end{equation}
where $\text{tgt}(r)$ denotes the target node type $\beta$ of the directional pair $r$, and $\mathbf{W}_{\text{root}}^{(r)} \in \mathbb{R}^{d_{\text{hidden}} \times d_{\text{hidden}}}$ is the trainable root projection header specific to that relation.

To stabilize gradient propagation and prevent representation oversmoothing across deep architectures, the final node hidden representation $\mathbf{h}_i^{(l)}$ is obtained by applying a non-linear activation function followed by a type-specific layer normalization operation:
\begin{equation}
\label{eq:final_gnn_layer}
\mathbf{h}_i^{(l)} = \text{LayerNorm}_{\beta} \left( \text{ReLU} \left( \mathbf{\tilde{h}}_i^{(l)} \right) \right), \quad \forall i \in \mathcal{V}_{\beta}.
\end{equation}
This layered sequence transforms local spatial topologies and raw physical channels into high-dimensional node-level feature vectors, establishing the spatial awareness required for global coordination.

\subsection{Global Attention Coordination Layer}
\label{subsec:global_attention}

While the relational message passing layers described in Section~\ref{subsec:gnn_message_passing} effectively encapsulate local topological layouts and link conditions, standard spatial convolutions suffer from a restricted receptive field governed by the structural neighborhood size. To address potential allocation dependencies over the network, individual reconfigurable intelligent surfaces can be coordinated globally to avoid redundant assignments where multiple surfaces serve overlapping physical areas. To overcome this limitation and enable unrestricted inter-surface information exchange, we project the final message-passed representations through a central Multi-Head Self-Attention (MHA) block~\cite{Vaswani2017}.

Let $\mathbf{H}_{\text{RIS}}^{(L)} \in \mathbb{R}^{|\mathcal{R}| \times d_{\text{hidden}}}$ denote the aggregated latent embedding matrix of all RIS nodes after the final message passing layer $l = L$. This matrix serves as the unified query, key, and value input to a non-local self-attention network operating over $H$ independent attention heads to compute collective spatial affinity scores. To preserve the localized structural qualities learned during spatial graph convolutions, we wrap this multi-head sequence in a residual shortcut connection. The final coordinated embedding matrix $\mathbf{\tilde{H}}_{\text{RIS}} \in \mathbb{R}^{|\mathcal{R}| \times d_{\text{hidden}}}$ is evaluated as:
\begin{equation}
\label{eq:attention_update}
\mathbf{\tilde{H}}_{\text{RIS}} = \text{LayerNorm}_{\text{att}} \left( \mathbf{H}_{\text{RIS}}^{(L)} + \text{MHA}\left( \mathbf{H}_{\text{RIS}}^{(L)} \right) \right),
\end{equation}
where $\text{MHA}(\cdot)$ denotes the integrated multi-head linear projection and scaled dot-product consensus mapping, and $\text{LayerNorm}_{\text{att}}$ denotes layer normalization applied to the attention residual output. This layer transforms the locally isolated structural vectors into network-aware features, providing the collective spatial insights required by the downstream assignment logic. Ultimately, while this multi-head self-attention block represents a complete, theoretically coordinated model, we utilize our downstream ablation studies to evaluate whether this global communication overhead is physically justified by propagation constraints.

\subsection{Actor-Critic Policy Heads}
\label{subsec:policy_heads}

The latent node-level representations refined by the graph convolutions and the non-local attention block are routed to two specialized architectural heads to evaluate the sequential allocations under an Advantage Actor-Critic learning framework. Rather than assigning fixed-dimension linear classification heads, which lack the capacity to process networks of varying sizes, we implement a parameter-sharing bilinear pairing network for the actor policy and a size-invariant global mean-pooling block for the critic network.

\subsubsection{Bilinear Assignment Head (Actor)}
The actor head generates a dynamic preference distribution over all physically viable candidate combinations between available RISs and base stations. Let $\mathbf{h}_r \in \mathbb{R}^{d_{\text{hidden}}}$ denote the final embedding vector for RIS node $r \in \mathcal{R}$, corresponding to the output of the final message-passing layer $\mathbf{h}_r^{(L)}$ from \eqref{eq:final_gnn_layer} (or the transpose of the $r$-th row of $\mathbf{\tilde{H}}_{\text{RIS}}$ in \eqref{eq:attention_update} when global attention is enabled)-and let $\mathbf{h}_b^{(L)} \in \mathbb{R}^{d_{\text{hidden}}}$ represent the final embedding vector for base station $b \in \mathcal{B}$ in \eqref{eq:final_gnn_layer}.

To calculate assignment preferences without depending on fixed node counts, the distinct representation matrices are projected into a shared $d_{\text{hidden}}$-dimensional latent space via independent, parallel multi-layer perceptrons ($\text{MLP}_{\text{RIS}}$ and $\text{MLP}_{\text{BS}}$):
\begin{align}
\boldsymbol{\phi}_r &= \text{MLP}_{\text{RIS}}\left( \mathbf{h}_r \right) \in \mathbb{R}^{d_{\text{hidden}}}, \quad \forall r \in \mathcal{R}, \\
\boldsymbol{\phi}_b &= \text{MLP}_{\text{BS}}\left( \mathbf{h}_b^{(L)} \right) \in \mathbb{R}^{d_{\text{hidden}}}, \quad \forall b \in \mathcal{B}.
\end{align}
The raw assignment preference logit $L_{r,b}$ tracking the structural matching suitability between surface $r$ and base station $b$ is evaluated directly via the inner product of their shared space projections:
\begin{equation}
\label{eq:bilinear_logits}
L_{r,b} = \boldsymbol{\phi}_r^T \cdot \boldsymbol{\phi}_b.
\end{equation}

This operation constructs a flexible preference matrix $\mathbf{L} \in \mathbb{R}^{|\mathcal{R}| \times |\mathcal{B}|}$. To restrict resource mappings strictly to the valid pathways outlined in Section~\ref{subsec:markov}, elements within $\mathbf{L}$ are set to $-\infty$ if surface $r$ has already been activated, or if no active communication link $(b,r)$ exists within the sparsified graph topology matrix. During training, the remaining unmasked components are normalized via a softmax transformation to compute the action selection distribution $\pi_\theta(a_t = (r,b) \mid s_t) \propto \exp(L_{r,b})$. Conversely, during deterministic deployment execution, the policy bypasses probabilistic sampling and directly executes the high-utility assignment maximizing the preference space via $a_t = \arg\max_{(r,b)} L_{r,b}$.

\subsubsection{State-Value Head (Critic)}
The centralized critic head estimates the expected cumulative return tracking the current state layout to guide policy parameter updates. To compress the global network layout into a unified context feature vector while maintaining invariance to both node permutations and scale expansions, we aggregate the node embeddings via mean-pooling across their respective subset populations:
\begin{equation}
\label{eq:pooling}
\mathbf{h}_{\text{global}} = \frac{1}{|\mathcal{R}|}\sum_{r \in \mathcal{R}} \mathbf{\tilde{h}}_r + \frac{1}{|\mathcal{B}|}\sum_{b \in \mathcal{B}} \mathbf{h}_b^{(L)}.
\end{equation}
This collective graph feature vector $\mathbf{h}_{\text{global}} \in \mathbb{R}^{d_{\text{hidden}}}$ encapsulates the joint macro-state of the active infrastructure layout. Finally, the centralized value tracking scalar $V_\psi(s_t) \in \mathbb{R}$ is evaluated by passing the pooled vector through a multi-layer regression perceptron ($\text{MLP}_{\text{value}}$):
\begin{equation}
\label{eq:value_head}
V_\psi(s_t) = \text{MLP}_{\text{value}}\left( \mathbf{h}_{\text{global}} \right),
\end{equation}
where $\text{MLP}_{\text{value}}$ contains a hidden linear layer parameterized by a Rectified Linear Unit ($\text{ReLU}$) activation and maps down to a single continuous output tracking state advantage baselines.

\subsection{Training Setup and Implementation Details}
\label{subsec:training_setup}
The proposed GNN-DRL framework is implemented using Python, PyTorch~\cite{Paszke2019}, and the PyTorch Geometric library~\cite{Fey2019}. Model training and empirical evaluations were executed on the TU Wien dataLAB high-performance computing cluster utilizing CUDA-enabled GPUs (predominantly NVIDIA L40S).

The reinforcement learning agent is trained for a maximum of $3 \times 10^6$ episodes, updating the policy gradients with a batch size of $32$ episodes. To ensure that a single policy network generalizes robustly across varying resource limits, the maximum active surface budget $K$ is sampled uniformly at random from $K \sim \mathcal{U}(1, 100)$ at the beginning of each training episode. To prevent overfitting and ensure reliable generalization across varying network geometries, we employ an early stopping mechanism. Specifically, the model is evaluated every $2,000$ training episodes against a fixed validation set of $50$ random layout seeds. If the average coverage probability on the validation set does not improve over a predetermined patience window, the training process is halted, and the network weights yielding the highest validation performance are retained for final testing.

\section{Complexity Analysis}
\label{sec:complexity}

In this section, we analyze the theoretical computational complexity of the proposed Graph Neural Network-driven Deep Reinforcement Learning framework and compare it against the iterative Greedy resource allocation benchmark. We formulate the computational overhead using Big-$\mathcal{O}$ asymptotic notation in terms of the number of base stations $|\mathcal{B}|$, reconfigurable intelligent surfaces $|\mathcal{R}|$, user equipments $|\mathcal{U}|$, and the allocation budget constraint $K \le |\mathcal{R}|$.

\subsection{Environmental State Extraction} 
Before any allocation decisions are made, the environment must extract the valid physical connections to form the candidate routing topology. A fully connected heterogeneous wireless network encompasses a worst-case edge complexity of $\mathcal{O}(|\mathcal{B}||\mathcal{R}| + |\mathcal{R}||\mathcal{U}| + |\mathcal{B}||\mathcal{U}|)$. However, the proposed environment applies an aggressive spatial consistency mask based on physical line-of-sight boundaries and path-loss thresholding. This geographic restriction dictates that a node's maximum connection degree is bounded by its local spatial density rather than the global network scale. Consequently, the number of active edges $|\mathcal{E}|$ grows linearly with the total number of nodes $|\mathcal{V}|$, meaning $|\mathcal{E}| \ll |\mathcal{B}||\mathcal{R}| + |\mathcal{R}||\mathcal{U}| + |\mathcal{B}||\mathcal{U}|$. Extracting this shared sparse graph state requires a localized initial complexity of $\mathcal{O}(|\mathcal{V}| + |\mathcal{E}|)$, which is utilized by both the baseline and proposed allocation frameworks.

\subsection{Complexity of the Greedy Baseline Framework}
\label{subsec:greedy_baseline_complexity}
The Greedy baseline algorithm operates by first randomly selecting a subset of $K$ unassigned RIS units. It then iteratively processes these chosen surfaces one by one. For each selected RIS, the algorithm evaluates potential assignments to its local visible candidate base stations. Because the greedy baseline leverages the pre-computed spatial graph constraints, the number of candidate base stations evaluated per RIS is bounded by a local density constant rather than the global network size $|\mathcal{B}|$.

However, during each trial assignment, the algorithm relies on a global metric evaluation to calculate the updated network coverage reward. This requires recomputing the cascaded channel matrices tracking the path gains from UEs to RISs, and RISs to BSs. Because this step utilizes dense matrix multiplications across the full network infrastructure, computing the updated state requires $\mathcal{O}(|\mathcal{U}| \cdot |\mathcal{R}| \cdot |\mathcal{B}|)$ operations per trial. 

Aggregating this metric evaluation across the localized candidate checks for all $K$ chosen surfaces, the total asymptotic complexity evaluates to:
\begin{equation}
    \mathcal{C}_{\text{Greedy}} = \mathcal{O}\left( K \cdot |\mathcal{U}| \cdot |\mathcal{R}| \cdot |\mathcal{B}| \right).
\end{equation}
Because the iterative metric evaluation inherently requires multi-dimensional dense array multiplications, the greedy framework exhibits a high-order polynomial scaling, posing severe computational bottlenecks for massive 6G layouts.

\subsection{Complexity of the Proposed GNN-DRL Framework}
Operating on the same sparsified graph state, the proposed GNN-DRL framework evaluates assignment decisions using a relational message-passing network optionally coordinated by a global self-attention layer.

\subsubsection{Heterogeneous Message Passing} 
The policy model utilizes $L$ layers of localized message passing via \texttt{WeightedEdgeConv} operators. Within a single layer, computing messages across active links requires $\mathcal{O}(|\mathcal{E}|)$ operations, while updating node features requires $\mathcal{O}(|\mathcal{V}|)$ operations, where $|\mathcal{V}| = |\mathcal{B}| + |\mathcal{R}| + |\mathcal{U}|$. For $L$ consecutive layers, the total computational complexity is stringently bounded by $\mathcal{O}(L(|\mathcal{V}| + |\mathcal{E}|))$.

\subsubsection{Self-Attention Coordination} 
To capture global submodular dependencies and avoid overlapping RIS assignments, the optional multi-head self-attention layer operates over the intermediate features of the RIS nodes. This mechanism computes an all-to-all alignment matrix exclusively among the surfaces, introducing a localized quadratic dependency of $\mathcal{O}(|\mathcal{R}|^2)$.

Because the DRL policy selects RIS-BS assignments sequentially over $K$ episodic steps, the combined total asymptotic complexity for the fully-featured GNN architecture (incorporating message passing and attention) is:
\begin{equation}
    \mathcal{C}_{\text{GNN-Attn}} = \mathcal{O}\left( K \cdot \left[ L(|\mathcal{V}| + |\mathcal{E}|) + |\mathcal{R}|^2 \right] \right).
\end{equation}

However, if the physical propagation environment dictates that global coordination is unnecessary, the self-attention layer can be completely bypassed. In this optimized, purely localized configuration, the total asymptotic complexity scales strictly linearly with the sparsified graph size:
\begin{equation}
    \mathcal{C}_{\text{GNN-Local}} = \mathcal{O}\left( K \cdot L(|\mathcal{V}| + |\mathcal{E}|) \right).
\end{equation}

To clearly understand the computational advantage of the neural approach, we compare the dominant non-linear scaling term of the attention-based framework, $K \cdot |\mathcal{R}|^2$, directly against the dense evaluation bottleneck of the greedy baseline, $K \cdot |\mathcal{U}| \cdot |\mathcal{R}| \cdot |\mathcal{B}|$. Factoring out the common linear terms $K \cdot |\mathcal{R}|$ from both expressions, this comparison reduces to evaluating the scale of the surfaces $|\mathcal{R}|$ against the product of active users and base stations $|\mathcal{U}| \cdot |\mathcal{B}|$. 

In dense urban deployments, the combined product of the number of users and base stations vastly exceeds the total number of deployed shared surfaces (i.e., $|\mathcal{R}| \ll |\mathcal{U}| \cdot |\mathcal{B}|$). Consequently, even the fully coordinated $\mathcal{C}_{\text{GNN-Attn}}$ configuration achieves a drastically lower computational overhead. Bypassing the attention block in the optimized $\mathcal{C}_{\text{GNN-Local}}$ configuration further removes the quadratic $|\mathcal{R}|^2$ dependency altogether, restricting the remaining runtime operations strictly to localized, linear sparse tensor manipulations that scale linearly with the network size. Crucially, both proposed GNN-DRL architectures execute entirely as highly parallelizable tensor operations natively accelerated by GPU hardware, representing a stark contrast to the iterative, CPU-bound sequential evaluations of the greedy baseline.

\section{Ablation Studies}
\label{sec:ablation_studies}
To systematically evaluate the architectural design choices of the proposed framework, we conduct ablation experiments across four core structural components: 1) the global self-attention mechanism, 2) the hidden feature dimension $d_{\text{hidden}}$, 3) the reward smoothing factor $\kappa$, and 4) the spatial neighborhood aggregation operators within the \texttt{WeightedEdgeConv} layers. All ablation configurations are evaluated across 3 independent training runs with different initialization seeds, with mean coverage probabilities and standard deviation bounds reported across varying active RIS activation budgets $K$.

\subsection{Impact of Global Self-Attention Coordination}
\label{subsec:ablation_attention}

\begin{figure}[t]
\centerline{\includegraphics[width=\columnwidth]{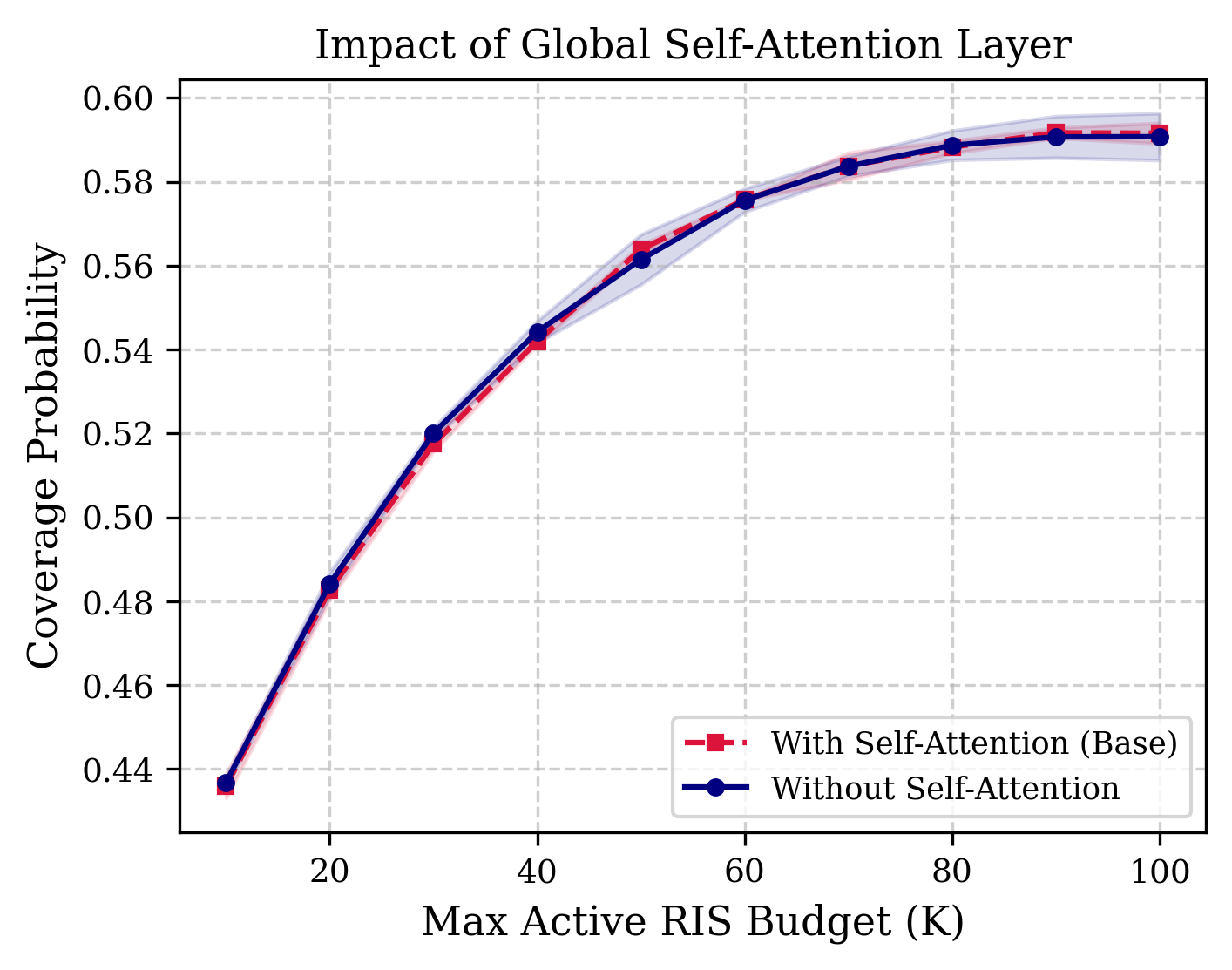}}
\caption{Coverage probability comparison between the model with global self-attention coordination and the localized graph convolutional architecture across varying active RIS activation budgets $K$.}
\label{fig:ablation_attention}
\end{figure}

In Section~\ref{subsec:global_attention}, we introduced an optional Multi-Head Self-Attention (MHA) block designed to evaluate inter-surface spatial affinity scores across all RIS nodes. However, incorporating global attention increases the computational complexity from linear $\mathcal{O}(K \cdot L(|\mathcal{V}| + |\mathcal{E}|))$ to quadratic $\mathcal{O}(K \cdot [L(|\mathcal{V}| + |\mathcal{E}|) + |\mathcal{R}|^2])$.

As shown in Fig.~\ref{fig:ablation_attention}, bypassing the self-attention layer yields practically identical coverage performance across all budget levels $K$. This empirical outcome confirms that multi-cell RIS resource dependencies are strictly localized by physical path loss. Coordination conflicts are confined to immediate geometric neighborhoods. Consequently, local spatial message passing in \texttt{WeightedEdgeConv} captures all necessary structural context, justifying the complete removal of the quadratic $\mathcal{O}(|\mathcal{R}|^2)$ attention block to maximize execution efficiency.

\subsection{Impact of Latent Feature Dimension $d_{\text{hidden}}$}
\label{subsec:ablation_hidden_dim}

\begin{figure}[t]
\centerline{\includegraphics[width=\columnwidth]{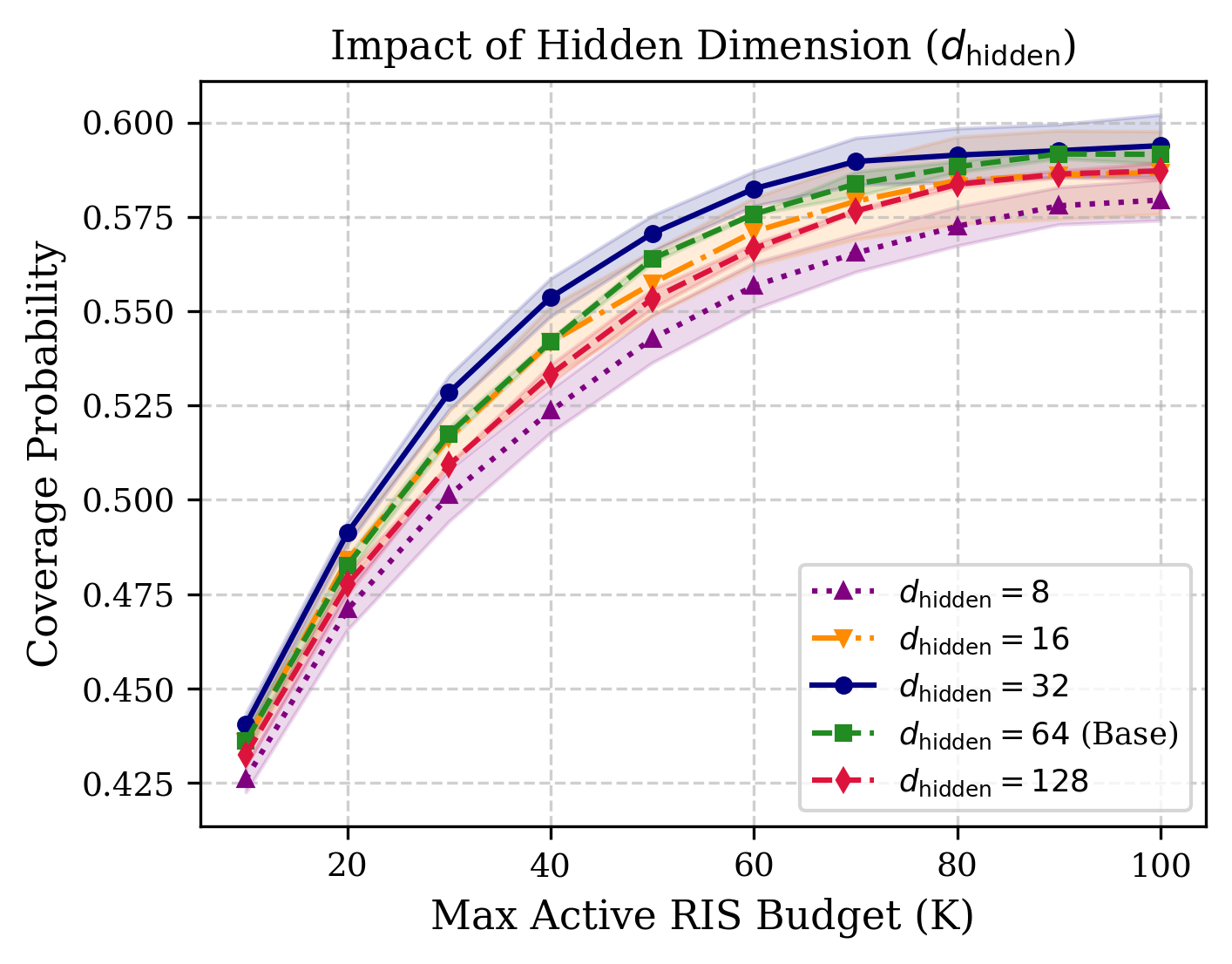}}
\caption{Impact of the hidden feature dimension $d_{\text{hidden}}$ on global coverage probability across varying active RIS activation budgets $K$.}
\label{fig:ablation_hidden_dim}
\end{figure}

The hidden feature dimension $d_{\text{hidden}}$ dictates the representational capacity of both the input linear projection headers and the relational message-passing layers. We evaluate $d_{\text{hidden}} \in \{8, 16, 32, 64, 128\}$.

As illustrated in Fig.~\ref{fig:ablation_hidden_dim}, $d_{\text{hidden}} = 32$ emerges as the clear optimal choice, consistently achieving the highest global coverage probability across all active RIS activation budgets $K$. A narrow representation space ($d_{\text{hidden}} = 8$) results in severe performance degradation due to underfitting, as an 8-dimensional latent vector lacks the capacity required to encode spatial node coordinates, infrastructure transmit powers, node degrees, and complex link attenuation metrics simultaneously. Increasing the dimension to $d_{\text{hidden}} = 16$ partially alleviates this bottleneck but remains noticeably suboptimal.

Conversely, expanding the capacity further to $d_{\text{hidden}} = 64$ and particularly $d_{\text{hidden}} = 128$ leads to a visible decline in performance compared to $d_{\text{hidden}} = 32$. In deep reinforcement learning settings, excessively large hidden dimensions introduce increased sample complexity. This higher capacity introduces greater variance into the advantage estimates during Actor-Critic policy updates, hindering smooth convergence and degrading final policy execution. Consequently, $d_{\text{hidden}} = 32$ establishes the ideal balance between model expressiveness, sample-efficient reinforcement learning, and low inference latency.

\subsection{Impact of Reward Smoothing Parameter $\kappa$}
\label{subsec:ablation_kappa}

\begin{figure}[t]
\centerline{\includegraphics[width=\columnwidth]{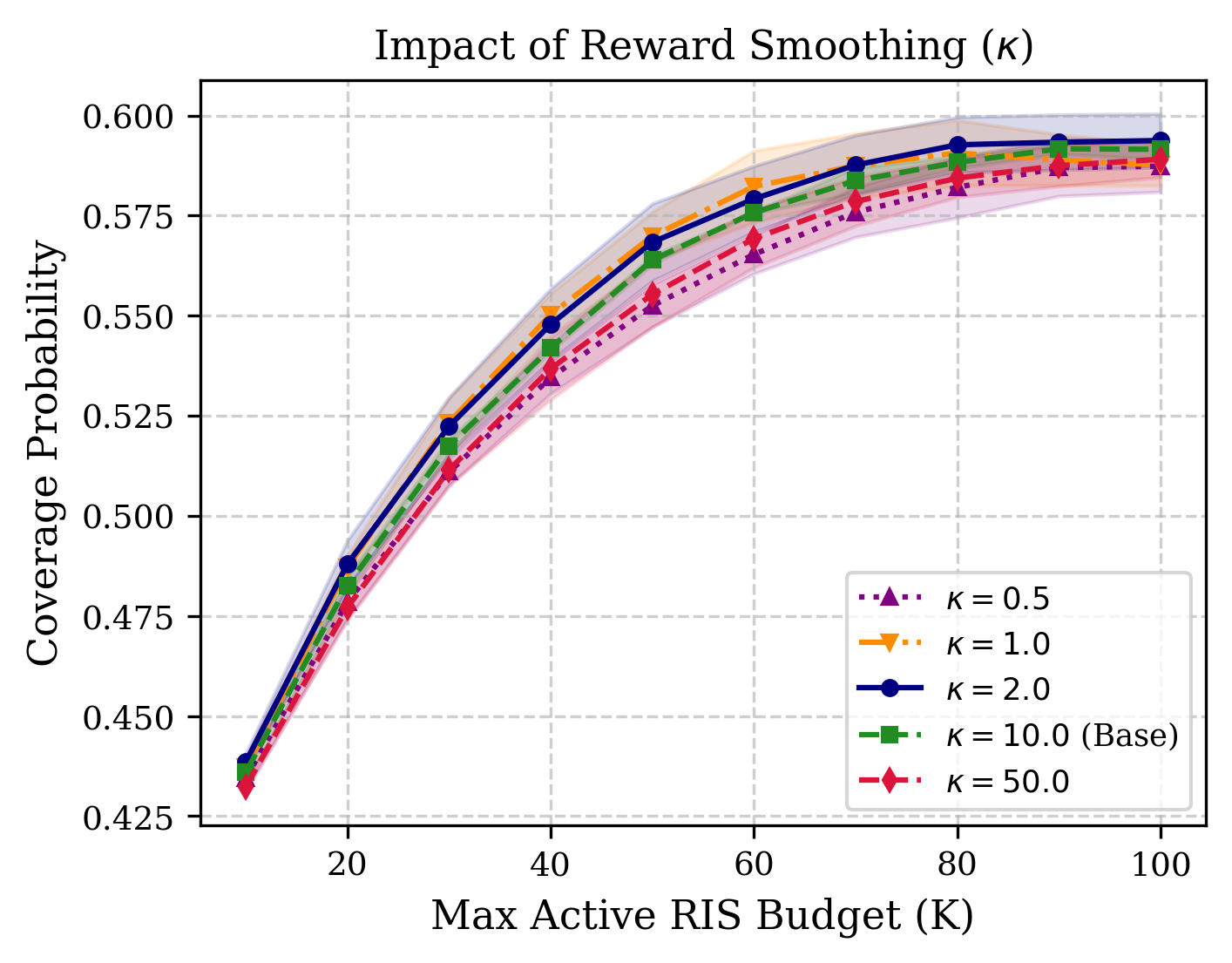}}
\caption{Impact of the reward steepness parameter $\kappa$ in the differentiable soft coverage surrogate \eqref{eq:soft_coverage} on policy convergence and coverage probability.}
\label{fig:ablation_kappa}
\end{figure}

To prevent zero-gradient plateaus during reinforcement learning, \eqref{eq:soft_coverage} utilizes a differentiable soft coverage surrogate controlled by the steepness parameter $\kappa$. We examine policy convergence for $\kappa \in \{0.5, 1.0, 2.0, 10.0, 50.0\}$.

As depicted in Fig.~\ref{fig:ablation_kappa}, setting $\kappa \in \{1.0, 2.0\}$ yields superior policy performance compared to both smaller and larger choices. When $\kappa = 0.5$, the surrogate function becomes overly flat, dampening the step-by-step marginal reward gains and slowing policy convergence. Conversely, excessively high values ($\kappa = 10.0$ and $\kappa = 50.0$) cause the sigmoid surrogate to approximate a sharp step function. This reintroduces gradient saturation near coverage decision boundaries, hindering advantage updates in the Actor-Critic framework. Therefore, setting $\kappa = 2.0$ provides the most effective continuous reward landscape for episodic training.

\subsection{Impact of Spatial Neighborhood Aggregation Operators}
\label{subsec:ablation_aggregation}

\begin{figure}[t]
\centerline{\includegraphics[width=\columnwidth]{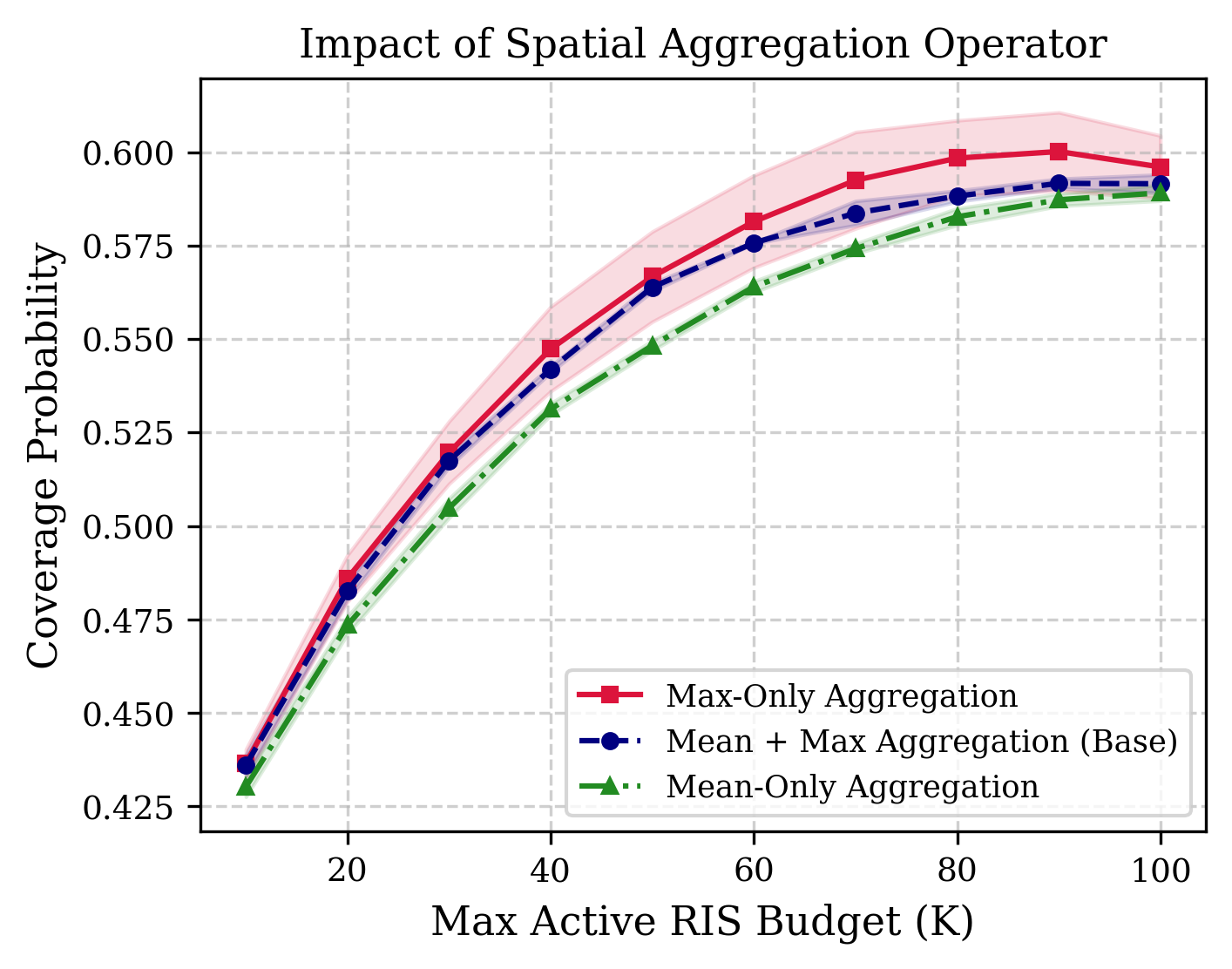}}
\caption{Coverage probability comparison among Mean-Only, Max-Only, and Dual (Mean + Max) spatial neighborhood aggregation operators within \texttt{WeightedEdgeConv}.}
\label{fig:ablation_aggregation}
\end{figure}

Within the \texttt{WeightedEdgeConv} layer \eqref{eq:aggregation_func}, messages from incoming neighbors are aggregated to update target node embeddings. We evaluate three spatial aggregation schemes: 1) Mean-Only aggregation, 2) Max-Only aggregation, and 3) Dual (Mean + Max) aggregation.

Fig.~\ref{fig:ablation_aggregation} compares the resulting coverage probabilities across active RIS budgets. Max-Only aggregation consistently achieves superior performance over both Mean-Only and Dual aggregation. From a physical perspective, wireless link quality and user SNR are dictated by the strongest available propagation path rather than the average quality of surrounding links. Mean aggregation tends to dilute high-gain direct or cascaded channels by averaging them with weaker background links. In contrast, Max-Only aggregation explicitly isolates peak local signal paths, enabling the GNN policy to identify optimal RIS-BS assignments more decisively. Furthermore, utilizing a single aggregation operator reduces the input dimensionality of the aggregation projection step in \eqref{eq:aggregation_func} from $2d_{\text{hidden}}$ to $d_{\text{hidden}}$, shrinking the trainable matrix $\mathbf{W}_{\text{aggr}}^{(r)}$ from $\mathbb{R}^{d_{\text{hidden}} \times 2d_{\text{hidden}}}$ to $\mathbb{R}^{d_{\text{hidden}} \times d_{\text{hidden}}}$ and further reducing the parameter footprint of the network.

\subsection{Impact of Physical Graph Sparsification}
\label{subsec:ablation_pruning}

\begin{figure}[t]
\centerline{\includegraphics[width=\columnwidth]{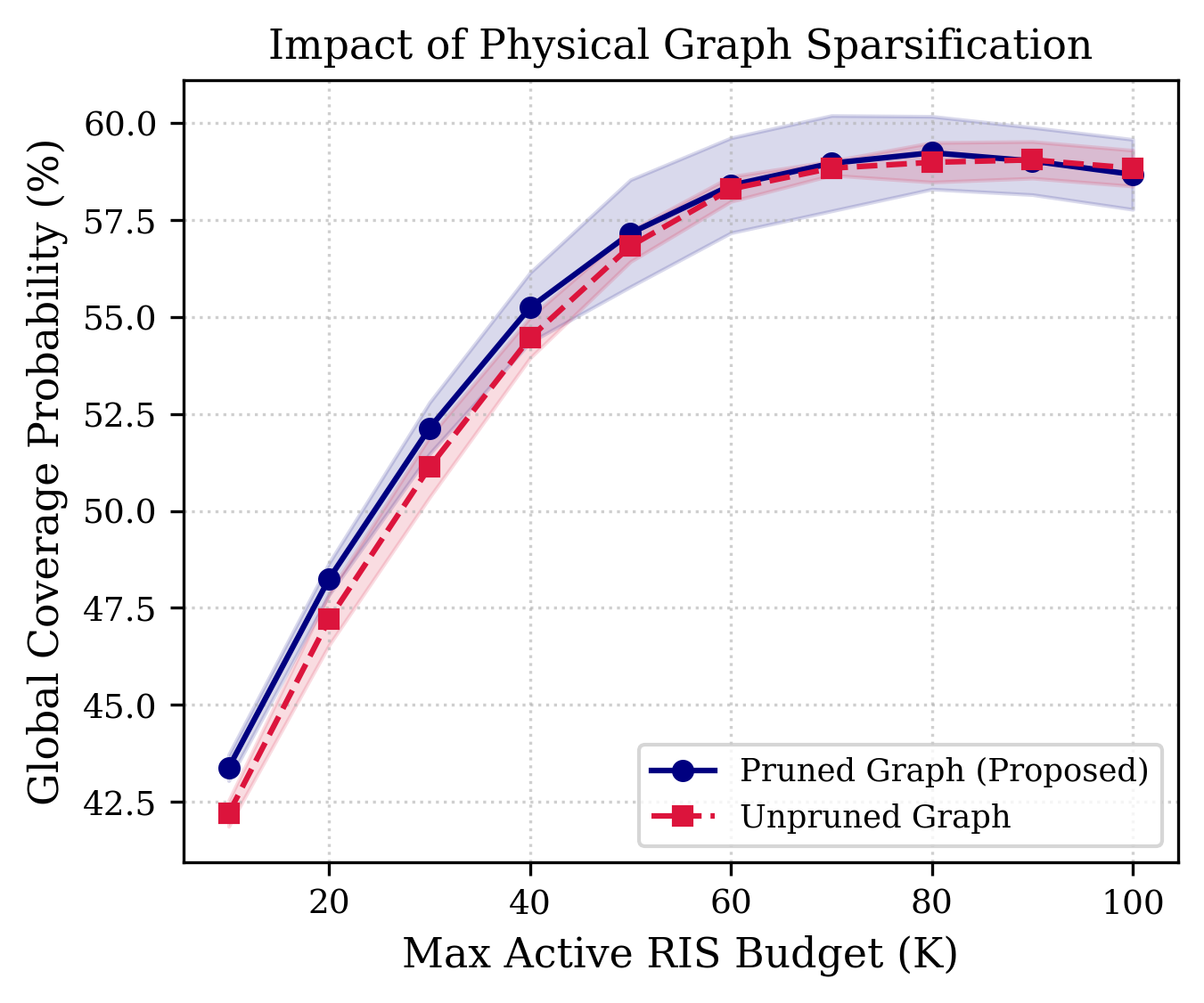}}
\caption{Global coverage probability comparison between the proposed sparsified physical graph and the unpruned baseline (all geometrically viable LOS links) across varying active RIS budgets $K$.}
\label{fig:ablation_pruning}
\end{figure}

To evaluate physical graph sparsification, we compare the proposed topology pruning strategy against an unpruned baseline graph retaining all geometrically viable LOS links across the network. For this study, the GNN policy utilizes the optimized architecture identified in previous ablations ($d_{\text{hidden}} = 32$, $\kappa = 2$, Max-Only aggregation, and no self-attention).

In a network layout featuring $8$ BSs, $\approx 100$ RISs, and $\approx 1000$ UEs, physical line-of-sight propagation naturally restricts unpruned RIS-UE candidate channels to $\approx 3,740$ links ($\approx 37$ visible UEs per surface). Applying performance-based path-loss thresholding ($\Delta_{\text{BU}} = 3\text{ dB}$, $\Delta_{\text{RU}} = 6\text{ dB}$) prunes weak non-serving connections, reducing BS-UE links by $\approx 84\%$ ($8,332 \to 1,313$) and RIS-UE links by $\approx 65\%$ ($3,740 \to 1,317$). Infrastructure BS-RIS links ($258$ active LOS paths) remain unaffected ($0\%$ pruned) due to their small initial footprint. Overall, graph sparsification compresses total edge density by $\approx 77\%$ ($12,331 \to 2,889$ edges).

As shown in Fig.~\ref{fig:ablation_pruning}, rather than degrading optimization accuracy, topology sparsification improves global coverage probability under constrained resource budgets ($K \le 60$). In dense, unpruned graphs, message passing aggregates feature embeddings across hundreds of weak, non-contributing channels, introducing representation noise. Physical topology pruning acts as an effective structural regularizer, enabling the GNN to focus message propagation strictly on high-utility candidate links.

\section{Numerical Results}
\label{sec:numerical_results}
To validate the scalability and performance of the proposed GNN-DRL framework, we conduct extensive simulations of a multi-cell mmWave network. The environment features spatially consistent blockages and highly clustered user distributions typical of dense urban layouts.

\subsection{Simulation Setup and Parameters}

The network is deployed over a $1000\text{ m}\times 1000 \text{ m}$ geographic area. A total of $8$ base stations are deployed according to a 2D Poisson disk process enforcing a minimum inter-site distance of $100\text{ m}$ to prevent unrealistic node co-location. User equipments ($\approx 1000$ UEs) follow a hybrid spatial distribution: $10\%$ are uniformly scattered across the area, while $90\%$ are concentrated in spatial user hotspots following a 2D Gaussian distribution with variance $\sigma_c^2 = 500\text{ m}^2$. Shared reconfigurable intelligent surfaces ($\approx 100$ units) are deployed directly at these hotspot centers to target coverage enhancement. 

Large-scale path loss and shadow fading parameters for direct and cascaded channels are modeled according to the 3GPP Urban Micro (UMi) scenario in 3GPP TR 38.901~\cite{3gpp38901}. Spatial blockage correlation is generated using a 2D spatially correlated Gaussian random field via FFT spectral shaping~\cite{Schwarz2016_LOS} with a decorrelation distance of $D_{\text{decorr}} = 50\text{ m}$. To reflect real-world hardware limits, each RIS comprises $M_{\text{RIS}} = 4096$ passive reflecting elements. Assuming standard half-wavelength ($\lambda/2$) element spacing at $f_c = 26\text{ GHz}$ ($\lambda \approx 1.15\text{ cm}$), a square array layout ($64 \times 64$) occupies a compact physical aperture of approximately $37\text{ cm} \times 37\text{ cm}$ ($\approx 0.14\text{ m}^2$), confirming its practical feasibility for mounting on building facades. The complete set of simulation parameters is summarized in Table~\ref{tab:sim_parameters}.

\begin{table}[t]
\renewcommand{\arraystretch}{1.2}
\caption{Simulation Parameters}
\label{tab:sim_parameters}
\centering
\begin{tabular}{l | l}
\hline\hline
\textbf{Parameter} & \textbf{Value} \\
\hline
Geographic Layout Area & $1000\text{ m} \times 1000\text{ m}$ \\
Number of Base Stations ($|\mathcal{B}|$) & $8$ \\
Number of Shared RISs ($|\mathcal{R}|$) & $\approx 100$ \\
Number of Users ($|\mathcal{U}|$) & $\approx 1000$ \\
Carrier Frequency ($f_c$) & $26\text{ GHz}$ \\
Subcarrier Bandwidth ($B$)  & $120\text{ kHz}$ \\
AWGN noise power spectral density ($N_0$) & $-174\text{ dBm/Hz}$ \\
Noise Figure ($F$) & $9\text{ dB}$ \\
BS Transmit Power ($P_b$) & $10\text{ dBm}$ \\
RIS Array Efficiency ($C_{\text{cluster}}$) & $1.0$ \\
RIS Elements ($M_{\text{RIS}}$) \& Aperture Size & $4096$ ($\approx 0.37\text{ m} \times 0.37\text{ m}$) \\
Target SNR Threshold ($\text{SNR}_{\min}$) & $0\text{ dB}$ \\
LOS Decay Distance: BS-UE ($d_{\text{LOS, BS-UE}}$) & $25\text{ m}$ \\
LOS Decay Distance: BS-RIS ($d_{\text{LOS, BS-RIS}}$) & $400\text{ m}$ \\
LOS Decay Distance: RIS-UE ($d_{\text{LOS, RIS-UE}}$) & $75\text{ m}$ \\
\hline\hline
\end{tabular}
\end{table}

\subsection{Baseline Algorithms}
To rigorously evaluate the allocation policy learned by the GNN, we compare its performance against four distinct baselines:
\begin{enumerate}
    \item \textbf{No RIS:} Represents the baseline macroscopic network performance relying exclusively on direct BS-UE propagation paths without any smart reflection assistance.
    \item \textbf{Random Assignment:} Iteratively selects up to $K$ unassigned RISs that possess at least one viable geometric line-of-sight to a base station, assigning each surface to a random visible BS candidate.
    \item \textbf{Greedy Optimization:} A sequential heuristic algorithm that iteratively selects RIS-BS assignments to maximize immediate global coverage gain. A detailed breakdown of its step-by-step operation and associated $\mathcal{O}(K \cdot |\mathcal{U}| \cdot |\mathcal{R}| \cdot |\mathcal{B}|)$ computational bottleneck is provided in Section~\ref{subsec:greedy_baseline_complexity}.
    \item \textbf{Idealistic Upper Bound:} A theoretical ceiling where the strict, unique-assignment constraint \eqref{eq:opt_unique} is relaxed. It assumes every valid RIS can simultaneously and coherently reflect signals for all users in the network with perfect phase alignment. While physically unachievable, it provides an absolute upper limit on cascaded propagation gains.
\end{enumerate}

\subsection{Spatial Allocation and Coverage Mapping}

\begin{figure}[t]
\centerline{\includegraphics[width=\columnwidth]{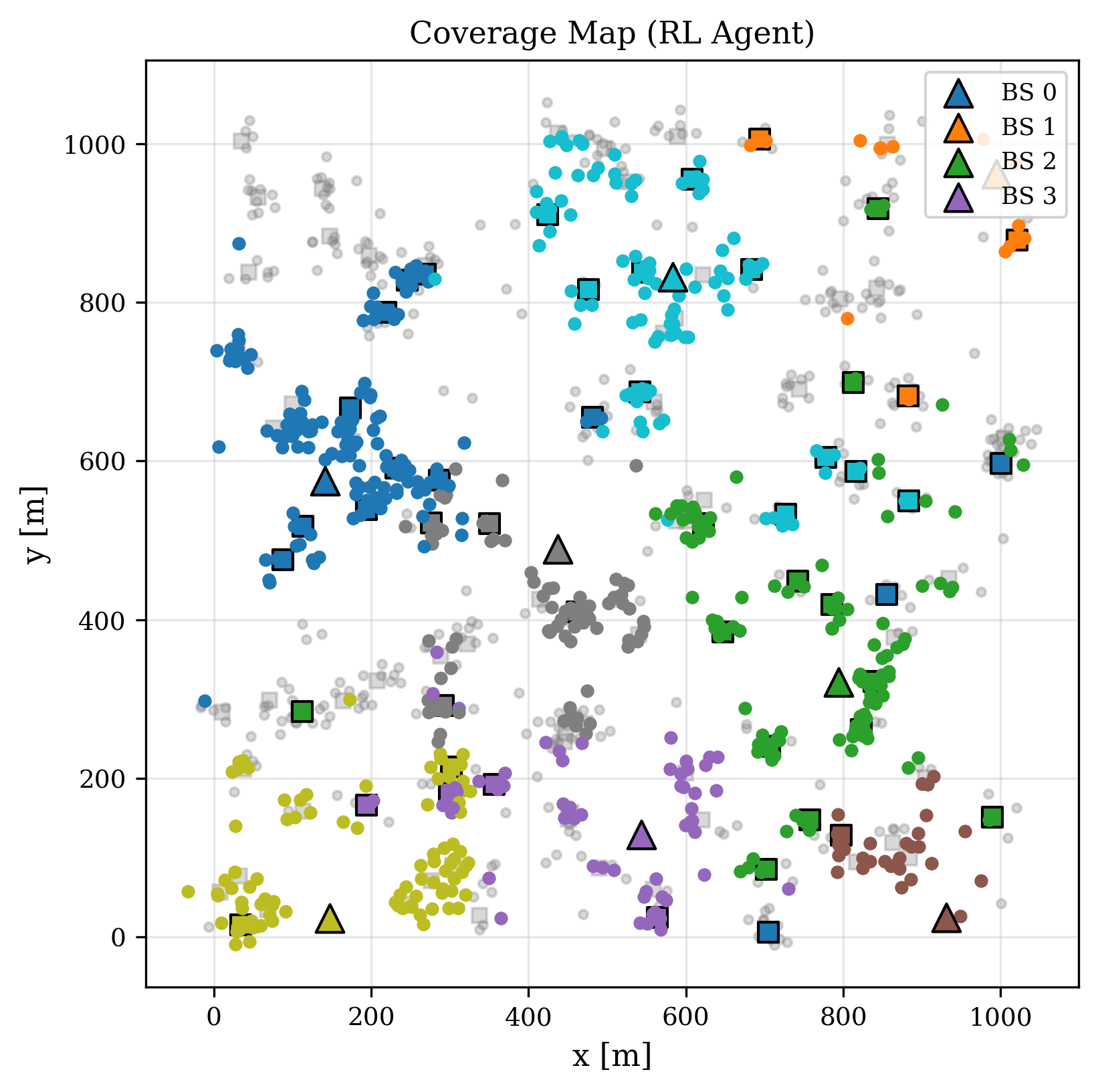}}
\caption{A representative multi-cell simulation snapshot illustrating the spatial deployment of base stations, shared RISs, and users, along with the dynamic RIS assignments generated by the trained GNN agent for an activation budget of $K = 50$.}
\label{fig:coverage_map}
\end{figure}

Fig.~\ref{fig:coverage_map} shows a representative spatial realization of the multi-cell environment along with the allocation matrix $\mathbf{X}$ determined by the GNN agent for an activation budget of $K = 50$. Base stations are represented by triangles, RIS units by squares, and users by small circles. 

Elements are color-coded to indicate cell association: a color match between an active RIS ($x_{r,b} = 1$) or a covered user ($\text{SNR}_u \ge \text{SNR}_{\min}$) and a base station indicates service by that base station. Unassigned RISs ($x_{r,b} = 0$) and unserved outage users ($\text{SNR}_u < \text{SNR}_{\min}$) are rendered as semi-transparent grey. Note that unserved users still associate with the base station providing the highest received power according to \eqref{eq:received_power}, but remain grey to highlight a coverage violation.

Interestingly, the policy occasionally assigns surfaces located at a significant 2D distance from their serving base stations. This behavior stems from the propagation environment: as established in Table~\ref{tab:sim_parameters}, the line-of-sight visibility range between base stations and elevated RIS units ($d_{\text{LOS, BS-RIS}} = 400\text{ m}$) is considerably larger than that for ground users ($d_{\text{LOS, BS-UE}} = 25\text{ m}$). As a result, distant base stations can establish strong cascaded links through unobstructed surfaces to assist shadowed user clusters. A detailed quantitative performance analysis across independent geometries is provided in the following subsections.

\subsection{Global SNR Enhancement}

\begin{figure}[t]
\centerline{\includegraphics[width=\columnwidth]{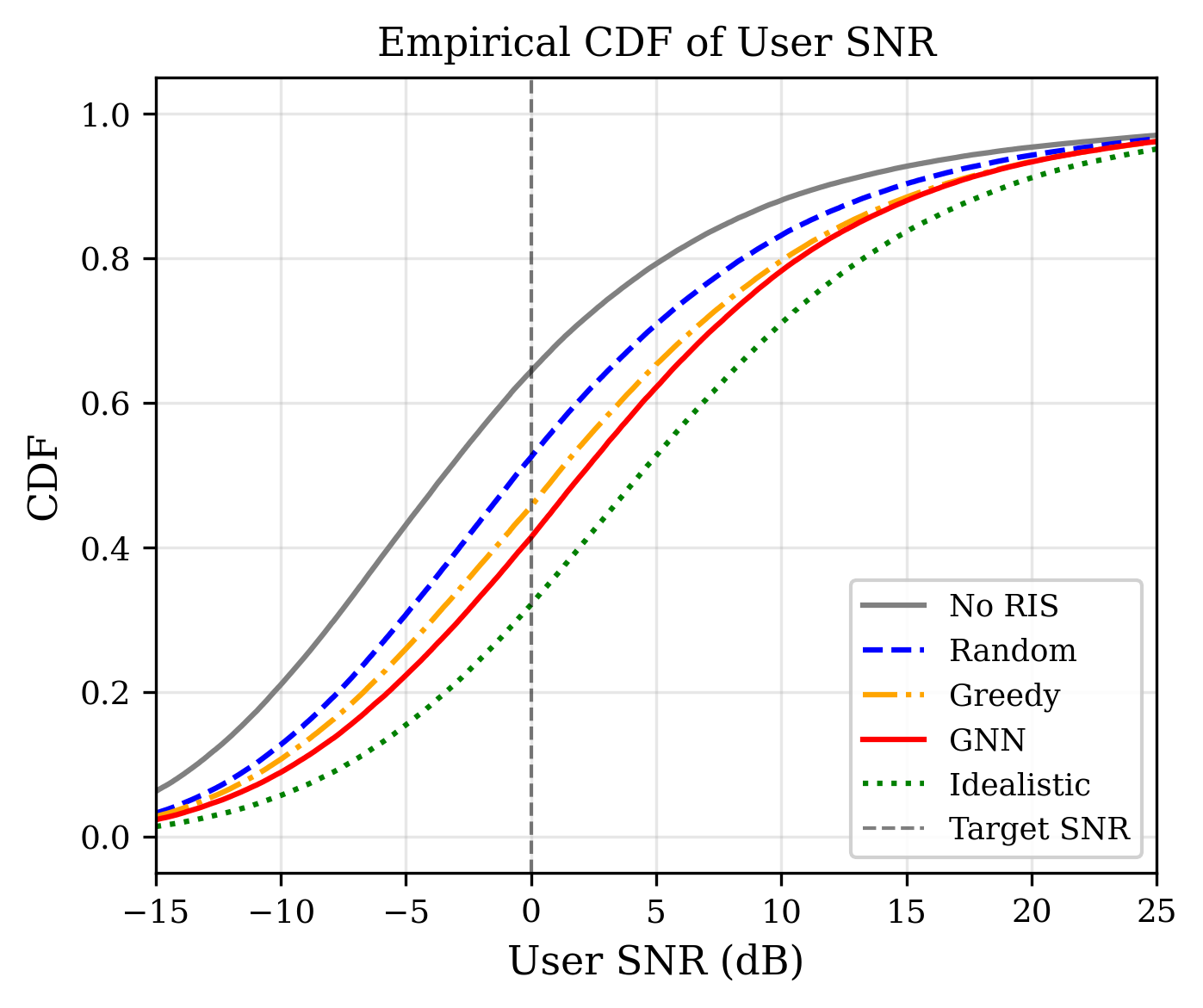}}
\caption{Empirical cumulative distribution function of user SNR across validation network realizations for an activation budget of $K = 50$.}
\label{fig:snr_ecdf}
\end{figure}

Fig.~\ref{fig:snr_ecdf} illustrates the empirical cumulative distribution function of user SNR across independent evaluation layouts for an active surface budget of $K = 50$. Deploying RIS infrastructure provides a noticeable SNR improvement across the network compared to the baseline without RIS assistance. While random surface assignment yields moderate gains, optimized association strategies perform substantially better.

Comparing the allocation methods, the proposed GNN policy achieves higher overall SNR levels than the sequential Greedy baseline. Interestingly, performance between the two optimized approaches is nearly identical at the extreme ends of the distribution: for the lowest $\sim5\%$ of users (severely shadowed cell-edge UEs) and the top $\sim15\%$ of users (UEs with strong direct LOS links), the GNN and Greedy allocations yield comparable results. The primary advantage of the GNN framework is concentrated across the middle percentile range, where it more effectively coordinates surface assignments to improve signal quality for moderately shadowed users.

\subsection{Impact of Shared Resource Constraints}

\begin{figure}[t]
\centerline{\includegraphics[width=\columnwidth]{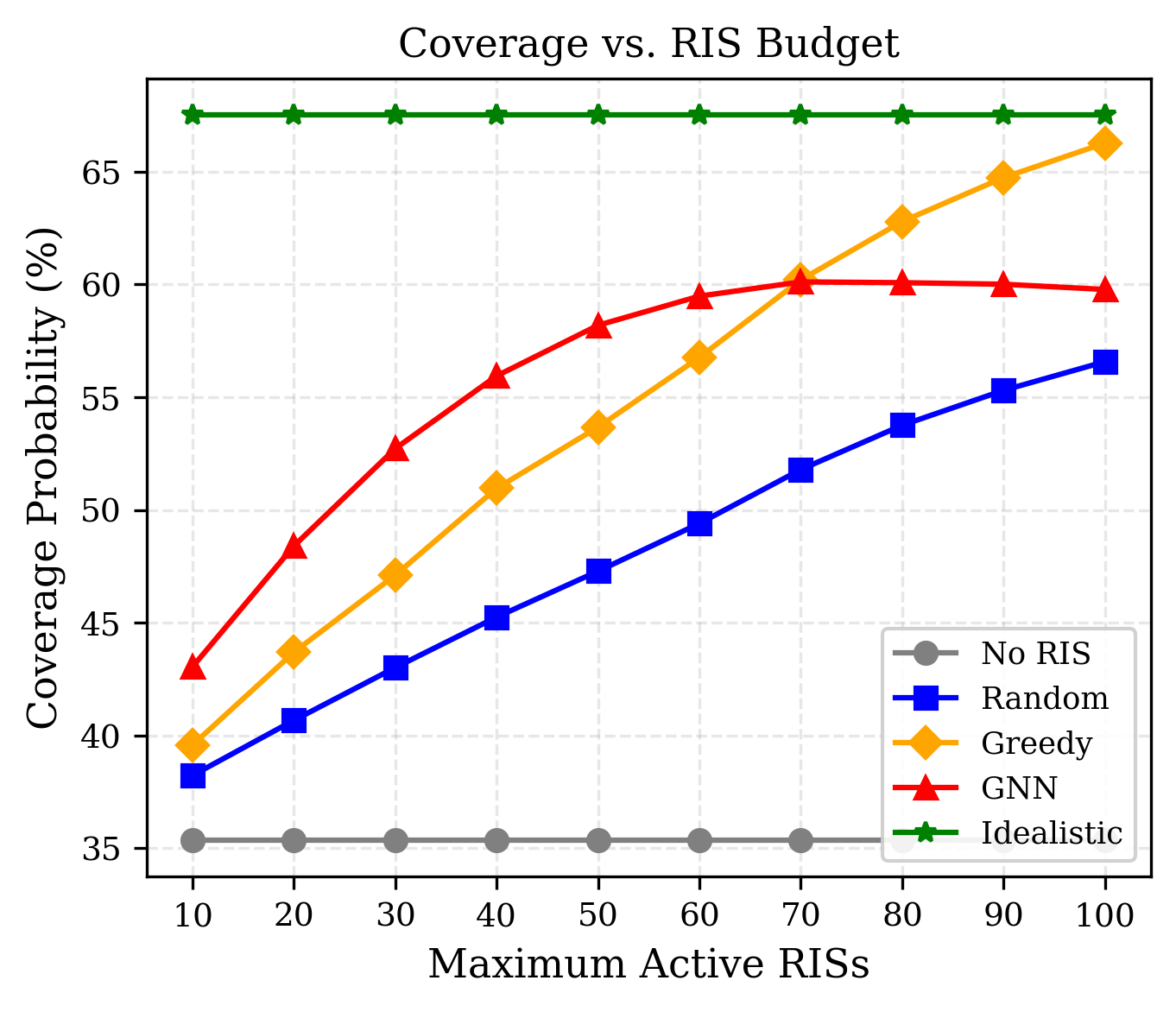}}
\caption{Global coverage probability as a function of the active surface activation budget $K$.}
\label{fig:budget_sweep}
\end{figure}

In practice, operational energy constraints or infrastructure licensing costs restrict the total number of active surfaces. To evaluate allocation efficiency under varying operational constraints, Fig.~\ref{fig:budget_sweep} plots the average coverage probability across an active surface budget sweep of $K \in [10, 100]$, evaluated using a single trained GNN policy without model retraining..

In resource-constrained regimes ($K \le 60$), the GNN policy consistently achieves superior coverage compared to the heuristic baselines. When resources are scarce, the budget constraint \eqref{eq:opt_budget} is strictly binding, making strategic surface selection critical. By leveraging multi-relational graph message passing across the local topology, the policy captures spatial synergies and long-term utility gains, avoiding the myopic choices of step-by-step greedy search. 

As the available budget expands, performance parity is reached around $K = 70$, beyond which the Greedy baseline gradually surpasses the GNN policy, becoming dominant as $K$ approaches 100. This shift occurs because as $K$ approaches the total surface count $|\mathcal{R}|$, the budget constraint \eqref{eq:opt_budget} becomes non-binding, effectively removing the combinatorial selection bottleneck as virtually all viable surfaces are activated. Furthermore, as network coverage approaches saturation near $K = 100$, the Greedy algorithm's exact step-by-step received power calculations allow it to exhaustively capture the remaining diminishing marginal coverage gains.

\section{Conclusion}
\label{sec:conclusion}

In this paper, we presented a scalable, heterogeneous Graph Neural Network-driven Deep Reinforcement Learning framework to solve the dynamic allocation of shared Reconfigurable Intelligent Surface infrastructure in dense 6G multi-cell networks. By formulating the NP-hard allocation problem as a sequential Markov Decision Process and leveraging physical line-of-sight viability alongside path-loss thresholding, we pruned dense multi-cell channel matrices into a highly sparse, heterogeneous tripartite graph. Coupled with our custom relational message-passing layer (\texttt{WeightedEdgeConv}), the Advantage Actor-Critic architecture natively generalizes to arbitrary layout dimensions and node counts without requiring model retraining.

Through rigorous asymptotic complexity analysis and comprehensive ablation studies, we proved that localized spatial message passing achieves strict linear computational scaling $\mathcal{O}(K \cdot L(|\mathcal{V}| + |\mathcal{E}|))$ relative to the sparsified graph topology. Crucially, our ablation results reveal that physical path loss localizes surface dependencies, rendering complex global self-attention mechanisms superfluous. Furthermore, our findings established that physical graph sparsification reduces overall edge density by 77\% and removes representation noise, directly enhancing global coverage probability while reducing computational complexity. In addition, we established that Max-Only spatial neighborhood aggregation consistently outperforms mean-based variants by preserving peak cascaded propagation links. 

Extensive numerical evaluations in dense urban mmWave environments validated the efficacy of the proposed approach. Under strict infrastructure resource constraints ($K \le 60$), the trained GNN agent consistently outperforms traditional greedy heuristics by up to approximately $12\%$ in global coverage probability, successfully mitigating the myopic decision bottlenecks of step-by-step search. Furthermore, the GNN policy executes via parallelizable tensor operations natively accelerated by GPU hardware, offering orders-of-magnitude faster inference speed compared to polynomial greedy channel computations.

Future work can extend this centralized GNN-DRL framework to highly dynamic multi-cell environments featuring continuous user mobility, time-varying spatial blockages, and real-time integrated sensing and communication tracking protocols. Additionally, investigating hierarchical multi-operator allocation frameworks-where dynamic inter-operator resource sharing protocols are combined with localized intra-operator GNN-DRL engines-presents a compelling direction for practical large-scale deployments.

\end{document}